# Towards Simultaneous Observation of Path and Interference of Single Photon in a Modified Mach-Zehnder Interferometer

Fenghua Qi, Zhiyuan Wang, Weiwang Xu, Xue-Wen Chen
School of Physics, Huazhong University of Science and Technology, Wuhan, 430074, China
*Corresponding author: E-mail:  xuewen_chen@hust.edu.cn

Zhi-Yuan Li
College of Physics and Optoelectronics, South China University of Technology, Guangzhou 510641, China
*Corresponding author: E-mail: phzyli@scut.edu.cn

[+]These two authors contributed equally to this work.

Abstract:
Classical wisdom of wave-particle duality says that it is impossible to observe simultaneously the wave and particle nature of microscopic object. Mathematically the principle requests that the interference visibility $V$ and which-path distinguishability $D$ satisfy an orthodox limit of $V^2+D^2 \leq 1$. This work presents a new wave-particle duality test experiment with single photon in a modified Mach-Zehnder interferometer and convincingly show the possibility of breaking the limit. The key element of the interferometer is a weakly-scattering total-internal reflection prism surface, which exhibits pronounced single-photon interference with a visibility up to 0.97 and simultaneously provides path distinguishability of 0.83. Apparently $V^2+D^2 \approx 1.63$ far exceeds the orthodox limit set by the principle of wave-particle duality for single photon. It is expected that more delicate experiments in future should be able to demonstrate the ultimate regime of $V^2+D^2$ approaching 2 and shed new light on the foundations of contemporary quantum mechanics.

# 1. Introduction

The wave-particle duality of microscopic particles, including photons, electrons, atoms, etc., constitutes the conceptual core of quantum physics. The principle dictates that all microscopic particles exhibit mutually exclusive behaviors of two intrinsic attributes of the wave nature and particle nature, namely, they behave either as wave or as particles, depending on how they are measured, but never both [1-9]. Over the past several decades, numerous studies using *gedanken* or practical interferometers like Mach-Zehnder interferometer (MZI) and Young's two-slit interferometer have been carried out to test the wave-particle duality of particles in various genuine arrangements, for instance, Wheeler's delayed-choice scheme [6,10-15], Scully's quantum eraser scheme [7-9], Afshar's scheme [16-18], and others [19-21], to name a few. The outcomes of all the previous test experiments can be summarized into three situations characterized by the interference visibility $V$ and which-path distinguishability $D$. In one situation, one use a specific setup to achieves perfect observation of the wave nature of particles with an interference visibility $V = 1$ at the price of completely losing the path distinguishability with $D = 0$. In the second situation, one applies another setup to distinguishes unanimously the path each particle passes at the cost of complete destruction of the interference pattern, i.e., $V = 0$ and $D = 1$. In the third situation, using interferometer setups with quantum delayed-choice schemes, one can observe simultaneous partial wave and partial particle nature of photons with $V \neq 0$ and $D \neq 0$ [12-15], yet still satisfying an orthodox limit of $V^2+D^2 \leq 1$ [22]. Despite tremendous efforts and an impressive progress, all the existing test experiments have shown no sign of going beyond the principle of wave-particle duality.

A rigorous analysis based on full quantum mechanics in Schrödinger picture reveals that previous interferometers, such as the delayed-choice scheme of MZI as well as Afshar's scheme, actually work in the strong-measurement regime, where the measurement of one entity (either wave or particle nature) will strongly disturb the other so that the observation of the two are exclusive [23,24]. As illustrated in Figure 1(a), a photon sent into the MZI goes either into path X or path Y with each 50% probability after the first beam splitter BS1. To know which-path the photon passes and thus the particle property, one needs to remove the second beam splitter BS2 and check the detectors X and Y. To observe the interference pattern and thus the wave property of photon, one should keep BS2 and examine the detectors X and Y. However, no matter how smart the design it is impossible to observe the wave and particle nature of photon simultaneously because the insertion and removal of BS2 are completely exclusive operations. Yet, as displayed in Figure 1(b), a modified MZI working in the weak-measurement regime, hereafter called WM-MZI, can largely get around the difficult mutual exclusion problem by using an interference screen to replace BS2 in the standard MZI [25]. This interference screen exhibits a high transmission and weak scattering of photon, and essentially becomes a detector of interference pattern via weak scattering. Similar idea has recently been extended to atom interferometers with weak-measurement path detectors [26]. Quantum mechanical analyses show that compared to standard MZI the proposed WM-MZI has the potential to make simultaneous observation of interference and which-path information of microscopic particles. Specifically, it can reach the level of $V = 1$ and $D \to 1$, so that $V^2 + D^2 \to 2$, which obviously far exceeds the regime allowed by the principle of wave-particle duality. In this work, we experimentally construct a WM-MZI setup and perform single-photon experiment with this new interferometer to test the wave-particle duality of photons.

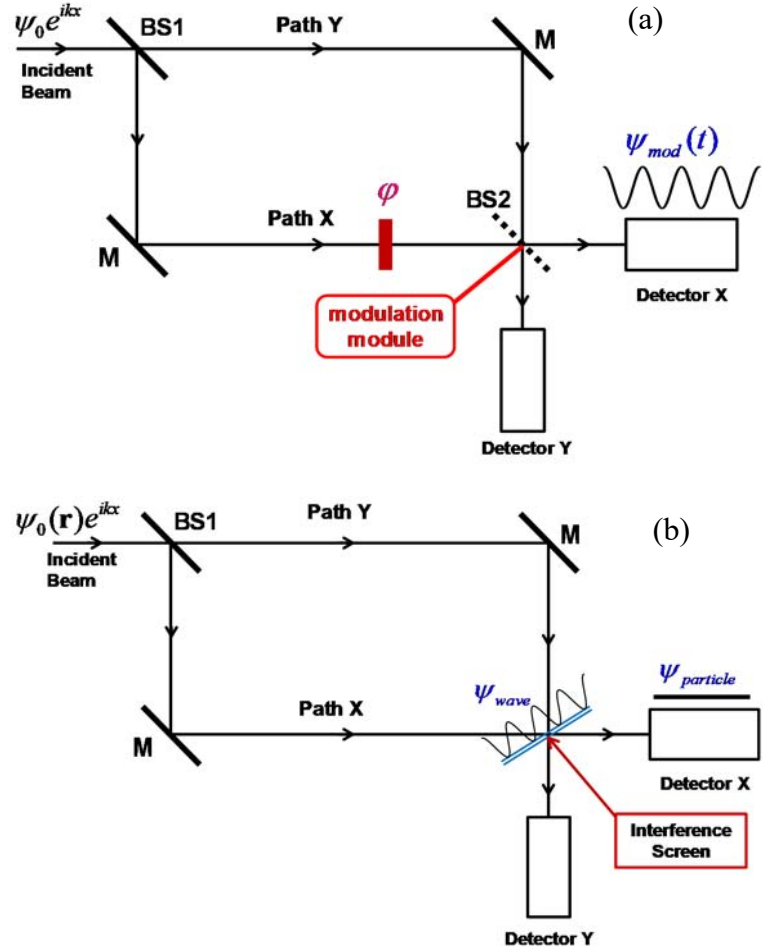

**Figure 1. (a)**. Schematic setup of classical Mach-Zehnder interferometer (MZI) used to test wave-particle duality of photon. The MZI consists of the first beam slitter BS1, two mirrors (M), a phase shift ($\varphi$), and the second beam splitter BS2. BS2 can either be present in the path, or absent, or controlled by an external (classical or quantum) module. **(b).** Schematic setup of weak-measurement MZI (WM-MZI), where an interference screen (denoted by the blue thick lines) with high transmission and weak scattering replaces BS2.

## 2. Construction of the WM-MZI and Experimental Setup

Simultaneous observation of the path and interference of single photon in the WM-MZI is an intriguing and practically challenging experiment. The difficulty lies in observing the interference pattern via the WM-MZI. The original experimental configuration in Figure 1(b) suffers from very low collection efficiency of weakly scattered single photons by the interference screen. For a two-dimensional image with 100×100 pixels, average photon detection rate at each pixel is estimated to be fewer than 0.01 count per second (cps) considering a single-photon source with an emission rate of 100 kcps, a scattering efficiency of 10%, and an overall detection efficiency of 1% for the proposed configuration. To circumvent such a formidable experimental difficulty, we take advantage of the fact that the interference pattern of two plane waves is one dimensional in nature and thus one can use a movable efficient point detector (an avalanche photodiode, APD) with a slit instead of a camera to observe the interference pattern. Moreover, we modify the original high-transmission weak-scattering interference screen in Figure 1(b) to a high-reflection weak-scattering prism setup so that a microscope objective with a high numerical aperture (NA) could be used to improve the collection of the weakly-scattered single photons. The above modifications afford us to experimentally demonstrate single-photon WM-MZI, as discussed in the following sections.

Figure 2(a) shows the key element of the WM-MZI, i.e. a prism surface that hosts the interference and provides weak scattering via a diffusive scattering thin film, which is a thin layer of milk on the prism surface. Two single-photon light beams respectively labeled as beam 1 and 2, are incident upon the prism surface with total internal reflections and an angle difference of $\theta \approx 1.75º$. Beam 1 and 2 exit the prism with the same separation angle and are detected by APD1 and APD2, respectively. The incident single-photon beams interfere on the prism surface to form a one-dimensional interference pattern, which normally evanescently decay into the air side of the surface. With a diffusive weak-scattering thin film, incident single photon is scattered by a small probability, a fraction of which is collected and detected to reveal the interference pattern. Moreover, with weak scattering the total reflection becomes imperfect, having a reflection coefficient $R$ slightly smaller than 1. We have tested the WM-MZI setup by incidence of a laser beam and observed very clear interference pattern at the prism hypotenuse surface, as illustrated in Figure 2(a), which confirms the excellent performance of the milk-coated prism as interference screen for light.

Figure 2(b) and 2(c) display the whole experimental setup. An efficient single photon source based on spontaneous emission of excited single CdSe/CdS coreshell colloidal quantum dots (QDs) [27-31] is prepared in Room 1 and delivered to the WM-MZI setup in Room 2 with a single mode fiber. The challenging part is to achieve large coupling of the single photons from the emission of single QDs into a single-mode fiber (SMF) [29]. Aiming for that goal, we design a sandwich sample structure as shown in the inset of Figure 2(b), where the QDs are sandwiched between the cover glass and a PMMA thin film with a thickness of 300 nm to ensure that a good amount of QD emission can be coupled to the SMF. The QD sample is excited by a 532 nm continuous-wave laser in an inverted microscope configuration and the fluorescent emission around 650 nm is filtered out and coupled into a SMF. We manage to have a coupling efficiency of about 15% so that the single-photon emission rate at the output of the SMF reaches about 100 kcps. As depicted in Figure 2(c), streams of single photons from the SMF are split into two arms via a 50:50 beam splitter and sent to the prism for the WM-MZI experiment as discussed previously. The optical path of one arm (path 2) is tuned by a wedge glass plate and a neutral density filter (NDF, 0.3 OD) in the other arm is used to balance the light intensity of the two arms. The weakly-scattered photons are collected by a microscope objective with a NA of 0.65 and sent to APD3 with a slit (150 μm width) for recording the interference pattern.

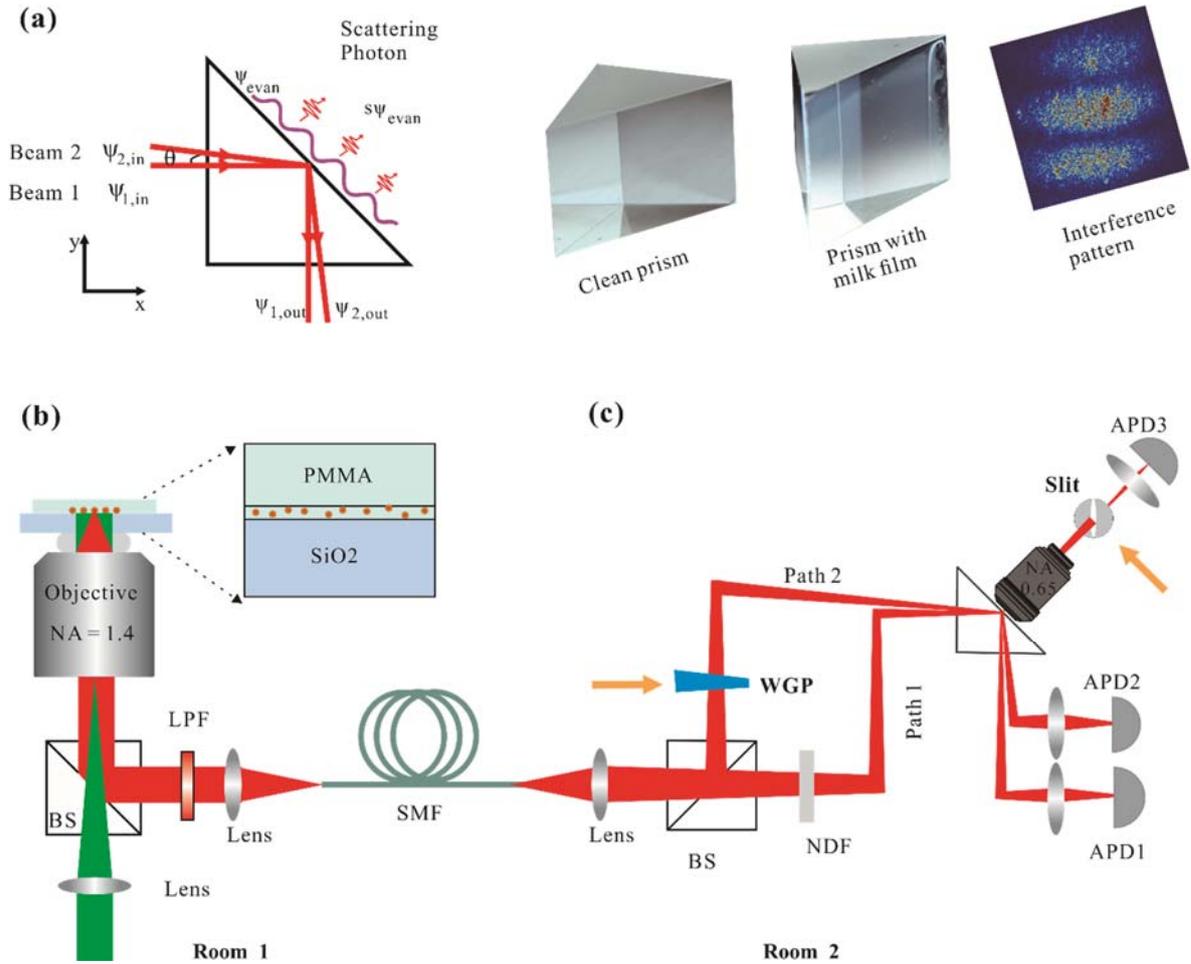

**Figure 2.** Schematics of the experimental setup of WM-MZI. **(a).** Photon interference on a prism surface coated with weakly scattering milk film as an interference screen. The interference pattern on the right is formed by the incidence of a laser beam into the WM-MZI setup. **(b).** A single-mode fiber (SMF) output single-photon source apparatus. Inset shows the sample structure where CdSe/CdS coreshell quantum dots (QDs) in PMMA served as single-photon emitters. LPF: longpass filter. **(c).** WM-MZI setup. The avalanche photon detectors APD1 and APD2 record two path way information, respectively. A wedge glass plate (WGP) is used to tune the optical length of Path 1. APD3 with a position-tunable slit is for observing the interference pattern.

Before launching the wave-particle duality test experiment, we characterize the setup and properties of the single-photon light source. We first measure the reflection of the prism surface coated with a weakly-scattering diffusive film by comparing the reflected light intensity from a silver mirror for the same source. The color-coded intensity time traces in Figure 3(a) and 3B show typical blinking phenomena of the photoluminescence (PL) from a single QD.

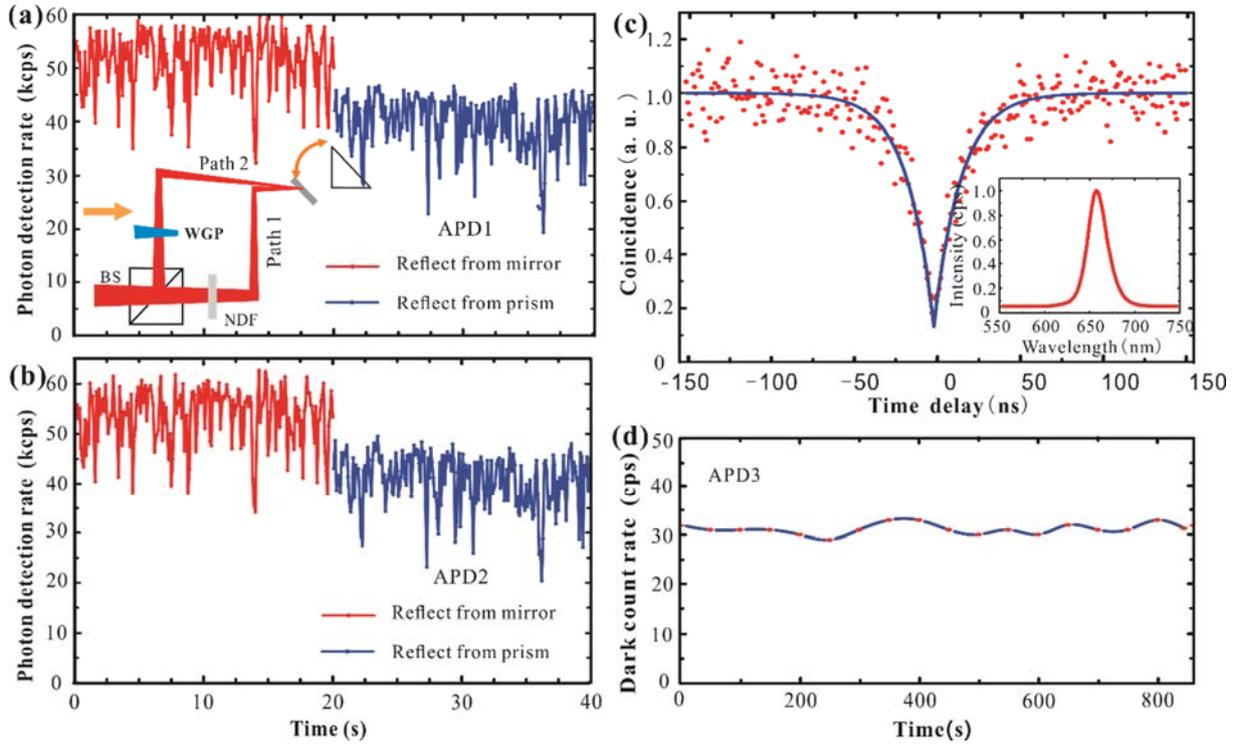

**Figure 3.** Characterization of the photon paths of the setup. **(a).** Time traces of the photon detection rates of APD1 for the cases of reflection from a silver mirror (red) and refection from a silver mirror (blue), respectively. Inset shows a schematic diagram of the measurement. **(b).** The same for APD2. **(c).** Coincidence measurement of the photon detection events of APD1 and APD2. Inset shows the spectrum of the photons in APD1 and APD2. **(d).** Dark count rate of APD3 as a function of time.

We decipher different charge states of the QDs (*31*) and only choose the bright state to evaluate the reflectivity. With the consideration of the reflectivity of silver mirror and two other surfaces of the prism, we estimate from the change of the average intensities of the two cases that the reflectivity of the prism coated with a weakly-scattering diffusive milk film is around 83.3%. Figure 3C displays the second-order photon correlation function g(2)(τ) and a pronounced anti-bunching dip at zero delay confirms single photon statistics of the source (27,30). A spectrum of the single-photon source is shown in the inset of Figure 3(c). Since APD3 will be used to detect the interference from a very weak signal, its dark count rate is characterized and shown as a function of time in Figure 3(d). An average dark account rate of 32 cps is measured and subtracted in the following measurements.

### 3. Theoretical Analysis of the WM-MZI

It is valuable to first make a quantum mechanical analysis of this WM-MZI and clarify its operation principle. The two incident single-photon beams are described by two wave functions

$$\psi_{1,in}(x,y) = \psi_0 e^{ikx+i\varphi}, \quad \psi_{2,in}(x,y) = \psi_0 e^{ik(x\cos\theta - y\sin\theta)} \quad (1).$$

Here $k = 2\pi/\lambda$ is the wavenumber of photon with wavelength λ and $\varphi$ is the phase shift of path 1 relative to path 2. The wave functions of the reflected beams read

$$\psi_{1,out}(x,y) = r\psi_0 e^{-iky+i\varphi}, \quad \psi_{2,out}(x,y) = r\psi_0 e^{ik(x\sin\theta - y\cos\theta)} \quad (2).$$

The reflection coefficient r is close to 1. The two beams interfere on the prism hypotenuse and the total field is evanescent into the air side of the surface. Its wave function can be written as

$$\begin{aligned}\psi_{evan}(x,y) &= t\psi_0 [e^{ikn\sin(45°+\theta_1)(x-y)/\sqrt{2}} + e^{ikn(x-y)/2+i\varphi}] \\ &\approx 2\psi_0 [e^{ikn(\sin\theta_1+\cos\theta_1)(x-y)/2} + e^{ikn(x-y)/2+i\varphi}]\end{aligned} \quad (3).$$

The scattered wave distribution recorded by APD3 can be given as

$$\psi_{scat}(x,y) = s\psi_{evan}(x,y) = 2s\psi_0 [e^{ikn(\sin\theta_1+\cos\theta_1)(x-y)/2} + e^{ikn(x-y)/2+i\varphi}] \quad (4).$$

Here $t = 1 + r \approx 2$ is the evanescent wave amplitude, $n$ is the refractive index of prism, $\theta_1$ is the refraction angle of single-photon beam within the prism, which satisfies Snell's law as $n\sin(\theta_1) = \sin(\theta)$. Mathematically, Eq. 1-4 are the solutions to Schrödinger equation for photon in this WM-MZI instrument. What do they mean? This deserves a careful quantum mechanical analysis and interpretation. From the orthodox quantum physics it follows that the wave function represents the spatial probability amplitude distribution of microscopic particle, including photon. In this framework, the two important physical quantities in Eq. 1-4, *r* and *s* represents the single-photon probability wave reflection and scattering coefficient, whereas $R = |r|^2$ and $S = |s|^2$ denote the reflection and scattering probability of single photon on the prism surface. Under the condition of small $S$ and large $R$, we find $S + R \approx 1$. Now from Eqs. 1-4, we obtain the single-photon light intensities of the two output paths as

$$P_{1,out} = |\psi_{1,out}(x,y)|^2 = RP_0, \quad P_{2,out} = |\psi_{2,out}(x,y)|^2 = RP_0. \tag{5}$$

where $P_0 = |\psi_0|^2$ is the input single-photon light intensity. The intensity distribution of the interference reads

$$P_{scat} = A|\psi_{scat}(x,y)|^2 = 8SAP_0\{1 + \cos[(kn\sin\theta_1 + kn\cos\theta_1 - kn)(x-y)/2 + \varphi]\}, \tag{6}$$

where $A$ is the combined collection and detection efficiency and is usually much smaller than 100%. In the above theoretical analysis, we have assumed a 100% transmission of photon through the two right-angle sides of the prism.

The next important step is to calculate the path distinguishability $D$ and the interference pattern fringe visibility $V$. Let us first make a brief analysis on what happens for a photon coming from beam 1 and 2. If the prism is not coated with a diffusive-scattering thin film, the total internal reflection is perfect, i.e. $r = 1, R = 1, s = 0$, so the photon will follow its own path and transport unanimously to photodetector APD1 and APD2, respectively. In this case the path distinguishability is 100%, while the interference pattern, although existing there at the prism hypotenuse outer surface, cannot be detected and thus the fringe visibility is zero. Yet, with the milk-coated prism as the interference screen, i.e., $r < 1, R < 1, s > 0$, and according to Eq. 6, it is possible to detect the interference pattern. Although the overall intensity $8SAP_0$ is low because of small value of $S$ and $A$, the fringe visibility $V$ theoretically can be as high as 1. The value of the path distinguishability can be estimated according to a simple model as follows. For each arm of WM-MZI, i.e., the reflection beam 1 and beam 2, if no scattering induced deviation of photon from its original path, the reflectivity should be ideally 100%. Now since scattering does occur, the extent of deviation of each photon from its original path is measured by the practical reflectivity $R$, thus the path distinguishability is taken simply as $D \approx R$.

## 4. Observations of the wave-particle test experiment in WM-MZI

In the following, we discuss the experimental results of the WM-MZI and clearly show that simultaneous observation of path and interference of single photons is achievable. As shown in the inset of Figure 4(a), the optical path length of Path 2 can be tuned by changing the position of the wedge plate (wedge angle of 0.5º) along the direction of the arrow. By tuning the wedge plate and using a fixed slit in front of APD3, we are able to measure the effect of longitudinal interference, i.e., interference with respect to phase shift $\varphi$, according to Eq. 6, by simultaneously counting single photons at three channels by using three APDs. The color-code traces in Figure 4(a) indicate the detected photon counts of reflected beam 1 and beam 2 for the bright states of the QD in blue and red, respectively. One observes that the photon counts levels of the two beams do not change much in particular for the wedge plate from 0.2 mm to 0.4 mm. The small variation is mainly due to the instability and blinking of the QD emission during the measurement time, which is quite normal. For APD3, the detected photon counts level is much lower and each data point is integrated for 50 seconds. Figure 4B displays the detected photon counts rate of APD3 (dark counts subtracted) as a function of the positon of the wedge plate.

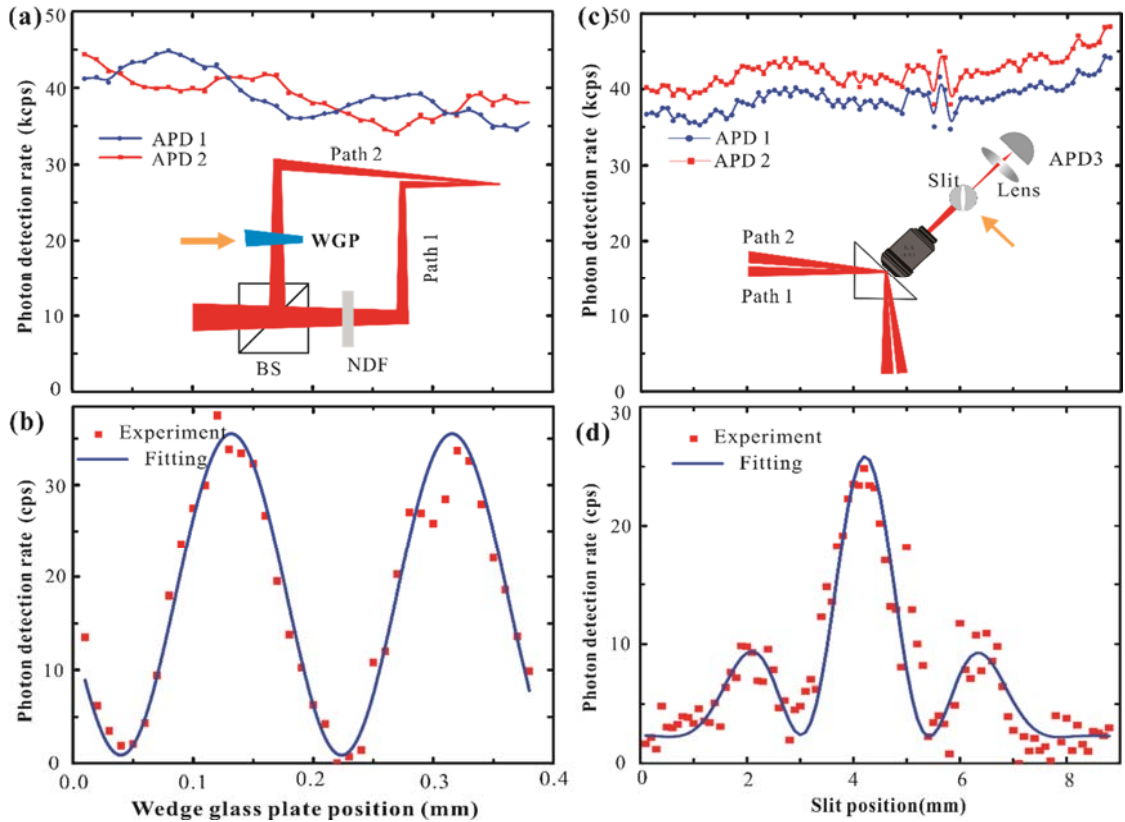

**Figure 4.** Simultaneous measurements of path and interference. For Longitudinal interference, **(a).** The photon detection rates of APD1 and APD2 change as a function of the optical path length tuned by the wedge glass plate (WGP). **(b).** The photon detection rate of APD3. The slit position is fixed in this series of measurements. For transverse interference, **(c).** the photon detection rates of APD1 and APD2 change as a function of the slit position. **(d).** the photon detection rates of APD3. The wedge glass plate is fixed in this series of measurements.

We remark that the time periods when the QD is in dark states emission have been subtracted and only the bright-state emission rates are shown in Figure 4 [31]. One observes a very nice periodic change of the signal with respect to the phase shift $\varphi$, which is the exciting longitudinal interference effect. The blue trace is a fit to the experimental data points and the visibility of the fringes $V$ is estimated to be 97%.

To directly observe the lateral interference pattern, i.e., the interference with respect to the displacement along the prism hypotenuse outer surface according to Eq. 6, we change the lateral position of the slit and again record the photon counts rate in these three channels simultaneously. In front of APD3, we use an aspherical condenser lens with a focal length of 16 mm to ensure that the whole spot where the two beams interfere can be imaged onto APD3 if the slit is removed. Therefore a lateral scan of the slit position offers the image on the spot. Figure 4(c) displays the detected photon counts of reflected beam 1 and beam 2 in blue and red, respectively. Again in principle, these channels are not affected with the change of the slit positon but in practice the signal levels fluctuate due to again the instable properties of colloidal QDs. Figure 4(d) shows the measured photon count rate of APD3 as a function of the slit position. One clearly observes the interference feature. On the other hand, the decrease of the fringe contrast with the lateral position difference is due to the finite spot size of the two beams. To evaluate the visibility of the lateral interference pattern, we model the interference with the assumption that two Gaussian beams with the same beam waists partially overlap in space with a small inclination angle. The blue trace in Figure 4(d) is a fit based on the model to the experimental results. We have estimated the visibility of the interference fringe $V$ to be 0.84 if the two beams were plane waves.

The above experimental data clearly indicates our WM-MZI exhibits an excellent performance for observing the wave property of single photon manifested from excellent interference pattern in both the longitudinal and lateral dimensions. The experiment agrees well with theoretical prediction as made in Eq. 6 and in Ref. [25]. Now we turn our eyes to another entity of wave-particle duality, i.e., the particle property of photon, which is connected with the path distinguishability $D$. In experiment, the reflection coefficient $R$ of the milk-coated prism is calibrated by comparing with a silver mirror. As shown in Figure 3(a), the photon signal is in random status with time elapsing, which means photons randomly emit one by one from a QD, pass through the WM-MZI, and detected by the photodetectors. The overall single-photon reflection intensity from the reference silver mirror and the milk-coated prism has a time-averaged quantity of 102 kcps and 81 kcps, respectively. Considering in practice the reflectivity of photon from the silver mirror is 95.5%, and the transmission coefficient of photon through each right-angle side of prism is 96%, we can calculate the reflection coefficient photon through the whole prism as *R*=83%, and the scattering loss is about 17%. This allows us to estimate the path distinguishability $D \approx R = 83\%$. As a result, we can make a good estimate to evaluate the wave-particle duality. For the longitudinal interference case we obtain $V^2 + D^2 \approx 0.97^2 + 0.83^2 = 1.63 \gg 1$ while for the lateral interference one we have $V^2 + D^2 \approx 0.84^2 + 0.83^2 = 1.39 \gg 1$.

The above experimental data using single photons agree well with the theoretical prediction made strictly in the framework of standard quantum mechanics formulation for the WM-MZI with a milk-coated prism surface as an interference screen. The weakly-scattering interference screen plays a key role in achieving a remarkable interference pattern, which in principle can exhibit a perfect fringe visibility as $V=1$ and in practice a very high value of $V=0.97$ is measured, despite that the absolute amplitude of this pattern is several orders of magnitude smaller than the incident single-photon light amplitude. At the same time, the two beams remain sufficiently high path distinguishability *D*. Thus this WM-MZI indeed can allow one to observe simultaneously the wave and particle feature of photon with a power much higher than that enabled by the well-established principle of wave-particle duality. We expect a better performance of obtaining $V = 1, D \to 1$ is achievable by improving the quality of single photon source, the diffusive milk-coating interference screen, the transmissivity through the right-angle side off prism, the collection and detection efficiencies and so on.

## 5. Conclusion

The above experimental observations obviously allow one to draw a basic physical picture about the whole journey a photon takes within the WM-MZI. A photon emitted from the QD source goes into the WM-MZI, passes the BS, goes to and transport along either path 1 or path 2, hits and goes into the prism horizontally, hit the hypotenuse surface and is reflected back downwards vertically, finally goes into the two photodetectors and triggers a signal count. This journey seems to be very ordinary, well known and understood, and nothing unexpected happens. However, when one looks closely at the hypotenuse surface of the prism, which is now coated with an everyday milk film, using some highly sensitive state-of-the-art single-photon detection and imaging instruments, something miracle happens. Right at the hypotenuse outer surface, very clear interference pattern forms when time goes on and more and more photons goes into the WM-MZI. A strange thing is that the fringe visibility is as perfect as comparable to the well-known case when an ordinary laser beam is used to perform the same experiments. In the language of quantum physics, the wave nature of photon is perfectly observed. The price to achieve this beautiful status is that the photon now does not follow strictly its original path of transport, however the deviation is only a little bit, and can be managed and reduced to a very low level when better instruments are used, so that one can be very sure (not 100% but close) of the path the photon eventually takes, or in the language of quantum physics, one can observe very good particle nature of photon. This is drastically different from the usual delayed-choice scheme of MZI as illustrated in Figure 1(a), where once perfect interference pattern is observed, the path each photon takes becomes completely ignorant to the observer.

Obviously the designed WM-MZI is much more powerful to observe simultaneous the wave and particle nature of photon than the delay-choice scheme of MZI. It should be emphasized that all these observations are dictated within the reign of standard quantum mechanics and can be predicted by standard calculations. Nonetheless, the physics underlying the unexpected behavior of photon is indeed going beyond the well-known principle of wave-particle duality. It seems our experiment does add some mysteries into the already mysterious system of concept upon orthodox quantum mechanics taught by the Copenhagen doctrine. We believe these experimental studies in a deeper and broader aspect in future will open up new insights upon the foundation of quantum physics and offer instructive clues as to explore more fundamental physics for microscopic world beyond contemporary quantum mechanics.

**Acknowledgements**

We gratefully acknowledge X. Peng, H. Qin and X. Hou (all of Zhejiang University, China) for providing the colloidal quantum dot sample and the continuous support. Funding: We acknowledge financial support from the National Natural Science Foundation of China (Grant Number 11874166,11434017, 11474114), Guangdong Innovative and Entrepreneurial Research Team Program (No. 2016ZT06C594), the Thousand-Young-Talent Program of China. F.Q. and Z. W. contributed equally to this work.

References

[1]. N. Bohr, in Albert Einstein: *Philosopher Scientist* (ed. P. A. Schilpp) 200-241 (Library of Living Philosophers, Evanston, 1949); reprinted in *Quantum Theory and Measurement* (eds. J. A. Wheeler and W. H. Zurek) 9-49 , Princeton Univ. Press, Princeton, **1983**.

[2]. R. Feynman, R. Leighton, and M. Sands, *The Feynman Lectures on Physics*, (Addison Wesley, Reading, **1965**, Vol. III, Ch. I.

[3]. W. K. Wootters and W. H. Zurek, *Phys. Rev. D* **1979**, *19*, 473.

[4]. M. Schlosshauer, *Rev. Mod. Phys.* **2004**, *76*, 1267.

[5]. A. D. Cronin, J. Schmiedmayer, and D. E. Pritchard, Rev. Mod. Phys. **2009**, *81*, 1051.

[6]. J. A. Wheeler, in *Mathematical Foundations of Quantum Theory*,(Eds: E. R. Marlow ) Academic Press, New York, **1978**, p. 9-48; J. A. Wheeler, in *Problems in the Formulations of Physics*, (Eds: G. T. di Francia), North-Holland, Amsterdam, **1979**.

[7]. M. O. Scully, B. G. Englert, and H. Walther, *Nature* **1991**, *351*, 111.

[8]. S. Dürr, T. Nonn, and G. Rempe, *Nature* **1998**, *395*, 33.

[9]. P. Bertet, S. Osnaghi, A. Rauschenbeutel, G. Nogues, A. Auffeves, M. Brune, J. M. Raimond, and S. Haroche, *Nature* **2001**, *411*, 166.

[10]. V. Jacques, E. Wu, F. Grosshans, F. Treussart, P. Grangier, A. Aspect, and J. F. Roch, *Science* **2007**, *315*, 966.

[11]. V. Jacques, E. Wu, F. Grosshans, F. Treussart, P. Grangier, A. Aspect, and J. F. Roch, *Phys. Rev. Lett.* 2008, **110**, 220402.

[12]. R. Ionicioiu and D. R. Terno, *Phys. Rev. Lett.* **2011**, *107*, 230406.

[13]. J. S. Tang, Y. L. Li, X. Y. Xu, G. Y. Xiang, C. F. Li, and G. C. Guo, *Nature Photon.* **2012**, *6,* 600.

[14]. F. Kaiser, T. Coudreau, P. Milman, D. B. Ostrowsky, and S. Tanzilli, *Science* **2012**, *338,* 637.

[15]. A. Peruzzo, P. Shadbolt, N. Brunner, S. Popescu, and J. L. O'Brien, *Science* **2012**, *338,* 634.

[16]. S. S. Afshar, E. Flores, K. F. McDonald, and E. Knoesel, *Found. Phys.* **2007,** *37,* 295.

[17]. O. Steuernagel, *Found. Phys*. **2007**, *37,* 1370.

[18]. V. Jacques, N. D. Lai, A. Dreau, D. Zheng, D. Chauvat, F. Treussart, P. Grangier, and J. F. Roch , *New J. Phys.* **2008**, *10*, 123009.

[19]. S. Kocsis, B. Braverman, S. Ravets, M. J. Stevens, R. P. Mirin, L. K. Shalm, A. M. Steinberg, *Science* **2011**, *332,* 1170.

[20]. A. Danan, D. Farfurnik, S. Bar-Ad, and L. Vaidman, *Phys. Rev. Lett.* **111**, 240402 (2013).

[21]. H. Yan, K. Liao, Z. Deng, J. He, Z. Y. Xue, Z. M. Zhang, and S. L. Zhu, *Phys. Rev. A* **2015**, *91*, 042132

[22]. B. G. Englert, *Phys. Rev. Lett.* 1996, **77**, 2154.

[23]. Z. Y. Li, *Chin. Phys. B* **2014**, *23*, 110309.

[24]. Z. Y. Li, *Chin. Phys. Lett.* **2016**, *33*, 080302.

[25]. Z. Y. Li, *Eur. Phys. Lett.* **2017**, *117*, 50005.

[26]. Z. Y. Li, Chin. Phys. B **2019**, *28*, 060301.

[27]. P. Michler, A. Kiraz, C. Becher, W. V. Schoenfeld, P. M. Petroff, L. Zhang, E. Hu, and A. Imamoglu, *Science* 2000, **290**, 2282.

[28]. H. Qin, Y. Niu, R. Meng, X. Lin, R. Lai, W. Fang, and X. Peng, *J. Am. Chem. Soc.* **2014**, *136,* 179.

[29]. K. G. Lee, X. W. Chen, H. Eghlidi, P. Kukura, R. Lettow, A. Renn, V. Sandoghdar, and S. Götzinger, *Nature Photonics* **2011**, *5*, 166.

[30]. G. Nair, J. Zhao, and M. G. Bawendi, *Nano Lett.* **2011**, *11*, 1136.

[31]. W. Xu, X. Hou, Y. Meng, R. Meng, Z. Wang, H. Qin, X. Peng, X. W. Chen, *Nano Letters* **2017,** *17,* 7487.